%% file: monniaux-lpar08-article.tex
\documentclass[a4paper]{article}
\usepackage{aeguill,amsfonts,latexsym,amssymb,stmaryrd}
\usepackage{amsmath}
\usepackage{graphicx}
\usepackage{float} % hyperref+algorithm badly interact otherwise
\usepackage[T1]{url}
\usepackage{hyperref}
\usepackage{algorithm,algorithmic}

\title{A Quantifier Elimination Algorithm for Linear Real Arithmetic}
\author{David Monniaux\\
VERIMAG%
\thanks{VERIMAG is a joint laboratory of CNRS, Universit\'e Joseph Fourier and Grenoble~INP.}}

\usepackage{amsthm}

\makeatletter
\newcommand{\tableaularge}{\footnotesize\setlength\tabcolsep{3\p@}}
\makeatother

\newcommand{\redi}{0.9}
\newcommand{\redii}{0.7}
\newcommand{\rediii}{0.6}

\begin{document}
\input{monniaux-lpar08}
\end{document}

%% file: monniaux-lpar08.tex
\newcommand{\bbQ}{\mathbb{Q}}
\newcommand{\bbR}{\mathbb{R}}
\newcommand{\parts}[1]{\mathcal{P}(#1)}
\newcommand{\sem}[1]{\llbracket #1 \rrbracket}

\newcommand{\eabs}{\epsilon_{\text{abs}}}
\newcommand{\erel}{\epsilon_{\text{rel}}}

\newcommand{\lfp}{\text{lfp~}}

\newcommand{\abstr}[1]{#1^\sharp}

\newcommand{\definedAs}{\stackrel{\text{def}}{=}}

\renewcommand{\topfraction}{1.00}
\renewcommand{\floatpagefraction}{1.00}
\renewcommand{\textfraction}{0.00}
\renewcommand{\dbltopfraction}{1.00}
\renewcommand{\dblfloatpagefraction}{1.00}

\maketitle

\begin{abstract}
We propose a new quantifier elimination algorithm for the theory of linear real arithmetic. This algorithm uses as subroutines satisfiability modulo this theory and polyhedral projection; there are good algorithms and implementations for both of these. The quantifier elimination algorithm presented in the paper is compared, on examples arising from program analysis problems and on random examples, to several other implementations, all of which cannot solve some of the examples that our algorithm solves easily.
\end{abstract}

\sloppy

\section{Introduction}
Consider a logic formula $F$, possibly with quantifiers, whose variables lay within a certain set $S$ and whose atomic predicates are relations over $S$. The models of this formula are assignments of values in $S$ for the free variables of $F$ such that $F$ evaluates to ``true''. \emph{Quantifier elimination} is the act of providing another formula $F'$, without quantifiers, such that $F$ and $F'$ are \emph{equivalent}, that is, have exactly the same models. For instance, $\forall x~(x \geq y \Rightarrow x \geq 3)$ is equivalent to quantifier-free $y \geq 3$.

If $F$ has no free variables, then $F'$ is a ground (quantifier-free, variable-free) formula. In most practical cases such formulas can be easily decided to be true or false; quantifier elimination thus provides a \emph{decision procedure} for quantified formulas.

In this paper, we only consider relations of the form $L(x, y, z, \dots) \geq 0$ where $L$ is a linear affine expression (an arithmetic expression where multiplication is allowed only by a constant factor), interpreted over the real numbers (or, equivalently, over the rationals). We can thus deal with any formula over linear equalities or inequalities. Our algorithm transforms any formula of the form $\exists x_1, \dots, x_n~F$, where $F$ has no quantifiers, into a quantifier-free formula $F'$ in disjunctive normal form. Nested quantifiers are dealt with by syntactic induction: in order to eliminate quantifiers from $\exists x~F$ or $\forall x~F$, where $F$ may contain quantifiers, one first eliminates quantifiers from~$F$. Universal quantifiers are converted to existential ones ($\forall x_1, \dots, x_n~F \equiv \neg \exists x_1, \dots, x_n~\neg F$), yet our algorithm generally avoids the combinatorial explosion over negations that hinders some other methods.

Our method can be understood as an improvement over the approach of converting to DNF through ALL-SAT and performing projection; we compared both approaches experimentally (see~\S~\ref{part:benchmarks}). We compared our implementation with commercial and noncommercial quantifier elimination procedures over some examples arising from practical program analysis cases, as well as random problems, and ours was the only one capable of processing them without exhausting memory or time, or failing altogether due to the impossibility of handling large coefficients.

\section{The Algorithm}
We first describe the datatypes on which our algorithm operates, then the off-the-shelf subroutines that it uses, then the algorithm and its correctness proof, then possible alterations.

\subsection{Generalities}
We operate on unquantified formulas built using $\wedge$, $\vee$, $\Rightarrow$, $\neg$ or other logical connectives such as exclusive-or (the exact set of connectives allowed depends on the satisfiability tester being used, see below; in this paper we shall only use $\wedge$, $\vee$ and $\neg$), and on quantified formulas built with the same connectives and the existential ($\exists$) and universal ($\forall$) quantifiers. It is possible to quantify not only on a single variable but also on a set of variables, represented as a vector $\vec{v}$. The atoms are linear inequalities, that is, formulas of the form $c + c_x x + c_y y + c_z z \dots \geq 0$ where $c \in \bbQ$ is the \emph{constant coefficient} and $c_v \in \bbQ$ is the coefficient associated with variable~$v$. It is trivially possible to represent equalities or strict inequalities using this formula language.
The \emph{models} of a formula $F$ are assignments $a$ of rational numbers to the free variables of $F$ such that $a$ satisfies $F$ (written $a \models F$). $F$ is said to be \emph{satisfiable} if a model exists for it. If $F$ has no free variables, then $F$ is said to be \emph{true} if $F$ is satisfiable, \emph{false} otherwise. Two formulas $A$ and $B$ are said to be \emph{equivalent}, noted $A \equiv B$, if they have the same models. Formula $A$ is said to \emph{imply} formula $B$, noted $A \Rrightarrow B$, if any model of $A$ is a model of $B$.

Consider a quantifier-free formula $F$, whose atomic predicates are linear inequalities, and variables $x_1,\dots,x_n$. We wish to obtain a quantifier-free formula $F'$ equivalent to $\exists x_1,\dots,x_n~F$. Let us temporarily forget about efficiency in order to convince ourselves quickly that quantifier elimination is possible. $F$ can be put into disjunctive normal form (DNF) $C_1 \vee \dots \vee C_m$ (by recursive application of distributivity), and $\exists x_1,\dots,x_n~F$ is thus equivalent to $(\exists x_1,\dots,x_n~C_1) \vee \dots \vee (\exists x_1,\dots,x_n~C_m)$. Various methods exist for finding a conjunction $C'_i$ equivalent to $\exists x_1,\dots,x_n~C_i$, among which we can cite Fourier-Motzkin elimination (see~\S~\ref{part:complexity}). We therefore obtain $F'$ in DNF. For a universal quantifier, through De Morgan's laws, we obtain a formula in conjunctive normal form (CNF).

Such a naive algorithm suffers from an obvious inefficiency, particularly if applied recursively to formulas with alternating quantifiers. A first and obvious step is to replace DNF conversion through distributivity by modern techniques (model enumeration using satisfiability modulo theory). We show in this paper than one can do better by interleaving the projection and the model numeration processes.

\subsection{Building blocks}
If one has propositional formulas with a large number of variables, one never converts formulas naively from CNF to DNF, but one uses techniques such as propositional satisfiability (SAT) solving. Even though SAT is NP-complete, there now exist algorithms and implementations that can deal efficiently with many large problems arising from program verification. In our case, we apply SAT modulo the theory of linear real inequalities (SMT), a problem for which there also exist algorithms, implementations, standard benchmarks and even a competition. Like SAT, SAT modulo linear inequalities is NP-complete. A SMT solver takes as an input a formula $F$ where the literals are linear equalities or inequalities, and answers either ``not satisfiable'', or a model of~$F$, assigning a rational number to each variable in~$F$. We assume we have such an algorithm \textsc{Smt} at our disposal as a building block

Another needed building block is quantifier elimination over conjunctions, named $\textsc{Project}(C, \vec{v})$: given a conjunction $C$ of linear inequalities over variables $\vec{v} = v_1, \dots, v_N$, obtain a conjunction $C'$ equivalent to $\exists v_1,\dots,v_n~C$. For efficiency reasons, it is better if $C'$ is minimal (no conjunct can be removed without adding more models), or at least ``small''. Fourier-Motzkin elimination is a simple algorithm, yet, when it eliminates a single variable, the output conjunction can have a quadratic number of conjuncts compared to the input conjunction, thus a pass of simplification is needed for practical efficiency; various algorithms have been proposed in that respect~\cite{imbert93fouriers}.
For our implementations, we used ``black box'' libraries implementing
geometrical transformations, in particular polyhedron projection: $C$ defines a convex polyhedron%
\footnote{A good bibliography on convex polyhedra and the associated algorithms can be found in the documentation of the Parma Polyhedra Library.~\cite{PPL} By \emph{convex polyhedron}, we mean, in a finite-dimension affine linear real space, an intersection of a finite number of half-spaces each delimited by a linear inequality, that is, the set of solutions of a finite system of linear inequalities. In particular, such a polyhedron can be unbounded. In the rest of the paper, the words ``polyhedron'' must be understood to mean ``convex polyhedron'' with that definition.} in $\bbQ^N$, and finding $C'$ amounts to computing the inequalities defining the projection of this polyhedron into~$\bbQ^{N-n}$.

\section{Quantifier Elimination Algorithm}
\label{part:algorithm}
We shall first describe subroutines \textsc{Generalize1} and \textsc{Generalize2}, then the main algorithm \textsc{ExistElim}.

\subsection{Generalized Models}

\begin{algorithm}
\caption{$\textsc{Generalize1}(a, F)$: Generalize a model $a$ of a formula $F$ to a conjunction}

\begin{algorithmic}
\REQUIRE{$a \models F$}
\STATE $M \gets \textsf{true}$
\FORALL {$P \in \textsc{AtomicPredicates}(F)$}
  \IF{$a \models P$}
    \STATE $M \gets M \wedge P$
  \ELSE
    \STATE $M \gets M \wedge \neg P$
  \ENDIF
\ENDFOR
\ENSURE{$M \Rrightarrow F$}
\end{algorithmic}
\end{algorithm}

\begin{algorithm}
\caption{\textsc{Generalize2}(G, M): Remove useless constraints from conjunction $M$ so that $G \wedge M \equiv \textsf{false}$}
\begin{algorithmic}
\REQUIRE{$G \wedge M$ is not satisfiable}
\FORALL{$c$ conjunct in $M$}
  \IF{$(G \setminus \{c\}) \wedge M$ is not satisfiable (call \textsc{Smt})}
     \STATE remove $c$ from $M$
  \ENDIF
\ENDFOR
\ENSURE{$G \wedge M$ is not satisfiable}
\end{algorithmic}
\end{algorithm}

Consider a satisfiable quantifier-free formula $F$. We suppose we have at our disposal a SMT-solving algorithm that will output a model $m \models F$. We wish to obtain instead a generalized model: a conjunction $C$ such that $C \implies F$. Ideally, we would like $C$ to have as few conjuncts as possible. We shall now see algorithms in order to obtain such generalized models.

The truth value of $F$ on an assignment $a$ of its variables only depends on the truth value of the atomic predicates of $F$ over $a$. Let us note $N_F = |\textsc{AtomicPredicates}(F)|$, where $|X|$ denotes the cardinality of the set $X$. These truth assignments therefore define at most $2^{N_F}$ equivalence classes over the valuations of the variables appearing in $F$. There can be fewer than $2^{N_F}$ equivalence classes, because some truth assignments can be contradictory (for instance, $x \geq 1$ assigned to \textsf{true} and $x \geq 0$ assigned to \textsf{false}). One can immediately generalize a model of a formula to its equivalence class, which motivates our algorithm \textsc{Generalize1}. Its output is a conjunction of literals from $F$.

\begin{figure}
\begin{center}
\includegraphics[scale=\rediii]{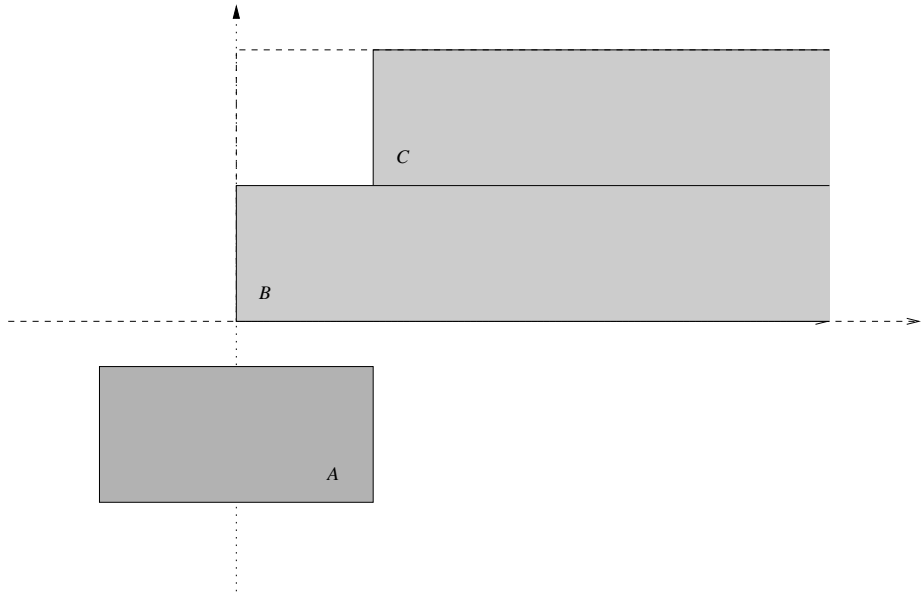}
\end{center}
\caption{$A$ is the first generalized model selected. If $G_0 \definedAs \neg F$, the initial value of $G$, is replaced at the next iteration by $G_1 \definedAs \neg F \wedge \neg \pi_0$ where $\pi_0$ is the projection of $A$, then it is possible to generate a single generalized model encompassing both $B$ and $C$ (for instance $x \geq -1 \allowbreak\wedge\allowbreak y \geq 0 \allowbreak\wedge\allowbreak y \leq 2$. If $G$ stays constant, then the $x \geq 1$ constraint defining the left edge of $C$ cannot be relaxed.}
\label{fig:generalize2}
\end{figure}

This conjunction may itself be insufficiently general. Consider the formula $F = (x \geq 0 \wedge y \geq 0) \vee (\neg {x \geq 0} \wedge y \geq 0)$. $x \mapsto 0, y \mapsto 0$ is a model of~$F$. \textsc{Generalize1} will output the conjunction $x \geq 0 \wedge y \geq 0$. Yet, the first conjunct could be safely removed. $\textsc{Generalize2}(\neg(F \vee O), M)$ will remove unnecessary conjuncts from $M$ while preserving the property that $M \Rrightarrow F \vee O$. Figure~\ref{fig:generalize2} illustrates why it is better to generalize the conjunctions.

The problem of obtaining a minimal (or at least, ``reasonably small'') inconsistent subset out of an inconsistent conjunction has already been studied. In DPLL(T) algorithms~\cite{Ganzingeretal2004CAV} for SMT-solving, the problem is to find out, given a consistent conjunction of literals $L_1 \wedge \dots \wedge L_n$ and a new literal $L'$, whether $L_1 \wedge \dots \wedge L_n \Rightarrow L'$, $L_1 \wedge \dots \wedge L_n \Rightarrow \neg L'$, or neither; and if one of the implications holds, produce a minimal \emph{explanation} why it holds, that is, a subset $L_{i_1}, \dots, L_{i_m}$ of the $L_i$ such that $L_{i_1} \wedge \dots \wedge L_{i_m} \Rightarrow L'$ (respectively, $\Rightarrow \neg L'$). Since this decision and explanation procedure is called often, it should be fast and much effort has been devoted in that respect by implementors of SMT-solvers (e.g. \cite{NieuwenhuisOliverasIC2007} for congruence theories). It is however not straightforward to use such explanation procedures for our purposes, since we do not consider conjunctions of literals only: when algorithm \textsc{ExistElim} invokes $\textsc{Generalize2}(\neg F, M_1)$, $\neg F$ is in general a complex formula, not a literal.

We therefore present here a straightforward inconsistent set minimization algorithm similar to the one found in~\cite[\S6]{dMRS:CADE2002}.
$\textsc{Generalize2}(G, M)$, where $M$ is a conjunction such that $G \wedge M$ is unsatisfiable, works as follows:
\begin{itemize}
\item
It attempts removing the first conjunct from $M$ (thus relaxing the $M$ constraint). If $G \wedge M$ stays unsatisfiable, the conjunct is removed. If it becomes satisfiable, then the conjunct is necessary and is kept.
\item The process is continued with the following conjuncts.
\end{itemize}

Unsurprisingly, the results of this process depend on the order of the conjuncts inside the conjunction $M$. Some orders may perform better than others; the resulting set of conjuncts is minimal with respect to inclusion, but not necessarily with respect to cardinality.
\footnote{%
This is the case even if we consider a purely propositional case. As an example, consider $F = A \vee (B \wedge C)$. $M = A \wedge B \wedge C \Rrightarrow F$, otherwise said $M \wedge \neg F$ is not satisfiable. If one first relaxes the constraint $A$, one gets the conjunction $B \wedge C$, which still implies~$F$; this conjunction has two propositional models ($A \wedge B \wedge C$ and $\neg A \wedge B \wedge C$). Yet, one could have chosen to relax $B$ and obtain $A \wedge C$, and then to relax $C$ and obtain $A$ (which still implies $F$); this formula has four propositional models.}

\begin{figure}
\begin{center}
\includegraphics[scale=\redi]{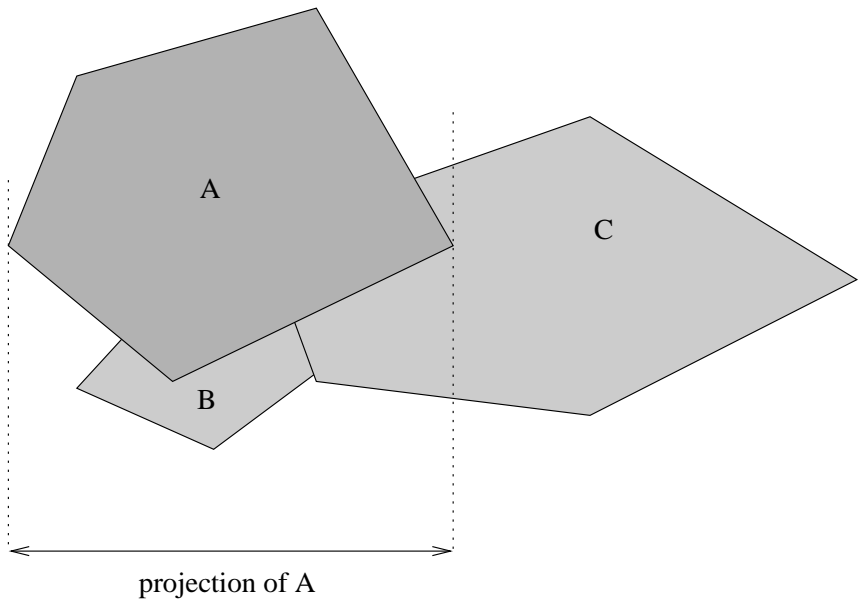}
\end{center}
\caption{Subsumption of one generalized model by another}
\label{fig:projection1}
\end{figure}

\begin{figure}
\begin{center}
\includegraphics[scale=\redii]{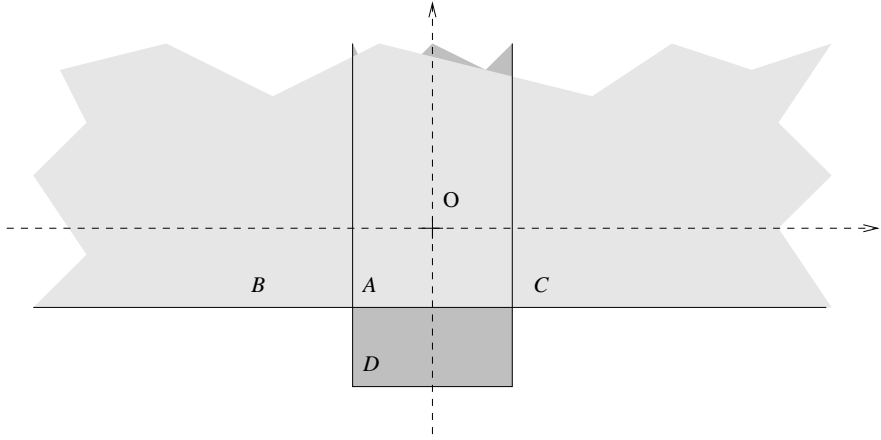}
\end{center}
\caption{The gray area is the set of points matched by formula $F = y \geq -1 \vee (y \geq -2 \wedge x \geq -1 \wedge x \leq 1)$. Point $\mathrm{O}=(0,0)$ is found as a model. This model is first generalized to $y \geq -1 \wedge y \geq -2 \wedge x \geq -1 \wedge x \leq 1$ according to its valuations on the atomic boolean formulas. Depending on whether one first tries to relax $x \geq -1$ or $y \geq -1$, one gets either a half plane (one conjunct) or a vertical band (three conjuncts); the former is ``simpler'' than the second. The simplicity of the formula output by \textsc{Generalize2} thus depends on the ordering of the input conjuncts.}
\label{fig:generalize1}
\end{figure}

\subsection{Main Algorithm}
\begin{algorithm}
\caption{\textsc{ExistElim}: Existential quantifier elimination}

\begin{algorithmic}
\STATE $H \gets F$
\STATE $O \gets \textsf{false}$
\WHILE[$(\exists \vec{v}~F) \equiv (O \vee \exists \vec{v}~H)$ and $H \wedge O \equiv \textsf{false}$ and $O$ does not mention variables from $\vec{v}$]{$H$ is satisfiable (call \textsc{Smt})}
  \STATE $a \gets$ a model of $H$ \COMMENT{$a \models H$}
  \STATE $M_1 \gets \textsc{Generalize1}(F, a)$ \COMMENT{$M_1 \Rrightarrow F$}
  \STATE $M_2 \gets \textsc{Generalize2}(\neg F, M_1)$ \COMMENT{$\neg(M_2 \wedge G)$}
  \STATE $\pi \gets \textsc{Project}(M_2, \vec{v})$
    \COMMENT{$\pi \equiv \exists \vec{v}~M_2$}
  \STATE $O \gets O \vee \pi$
  \STATE $H \gets H \wedge \neg \pi$
\ENDWHILE
\ENSURE{$O \equiv \exists \vec{v}~F$}
\end{algorithmic}
\end{algorithm}

The main algorithm is $\textsc{ExistElim}(F, \vec{v})$ which computes a DNF formula equivalent to $\exists \vec{v}~F$. $\vec{v}$ is a vector of variables. $\vec{v}$ can be empty, and then the algorithm simply computes a ``simple'' DNF form for~$F$. The algorithm computes generalized models of $F$ and projects them one by one, until exhaustion. It maintains three formulas $H$ and $O$. $O$~is a DNF formula containing the projections of the models processed so far. $H$ contains the models yet to be processed; it is initially equal to~$F$. For each generalized model $M$, its projection $\pi$ is added to $O$ and removed from~$H$.
$\textsc{ExistElim}$ can thus be understood as an ALL-SAT implementation coupled with a projection, where the projection is performed inside the loop so as to simplify the problem (as opposed to waiting for all models to be output and projecting them).

The partial correctness of the algorithm ensues from the loop condition and the following loop invariants:
$(\exists \vec{v}~F) \equiv O \vee (\exists \vec{v}~H)$,
$H \Rrightarrow F$ and 
$O$ does not mention variables from $\vec{v}$.

Given a formula $\phi$, we denote by $W(\phi)$ the number of equivalence classes induced by the atomic predicates of $F$ with nonempty intersection with the models of~$\phi$. Termination is ensured because $W(H)$ decreases by at least one at each iteration: $M_1$ defines exactly one equivalence class, $M_2$ defines a union of equivalence classes which includes the one defined by $M_1$, and the models of $\pi$ include those of $M_2$ thus also at least one equivalence class. The number of iterations is thus at most $2^{N_F}$. Note that \textsc{Generalize2} is needed neither for correctness nor for termination, but only for efficiency: otherwise, the number of iterations would always be the number of equivalence classes, which can be huge.

\section{Possible Changes and Extensions}
We investigated two variations of the same algorithm, both of which perform significantly worse. In addition, we extended the algorithm to quantifier elimination modulo a user-specified theory.

\label{part:variants}

\subsection{ALL-SAT then project  (Mod1)}
\label{part:algo-mod1}
The algorithm would still be correct if $M$ was removed from $H$ instead of $\pi$. It then becomes equivalent to performing ALL-SAT (obtaining all satisfying assignments) then projection. On the one hand, with this modified algorithm, the set of atomic formulas of $H$ would stay included in that of $F$ throughout the iterations, while this set can grow larger with the original algorithm since the set of atomic formulas of the projection of $F$ can be much larger than the set of atomic formulas in~$F$ (see~\S\ref{part:complexity}). On the other hand, the original algorithm may need fewer iterations because $\pi$ may subsume several generalized models, as shown by Fig.~\ref{fig:projection1}~: $A$ is the first generalized model being generated, and its projection subsumes~$B$; thus, the original algorithm will not have to generate $B$, while the modified algorithm will generate~$B$. Our experiments (\S\ref{part:benchmarks}) showed that the unmodified algorithm often performs much better in practice than this approach.

\subsection{Removals from Negated Set  (Mod2)}
\begin{algorithm}
\caption{\textsc{ExistElim}(Mod2): Existential quantifier elimination}
\small
\begin{algorithmic}
\STATE $H \gets F$
\STATE $G \gets \neg F$
\STATE $O \gets \textsf{false}$
\WHILE[$(\exists \vec{v}~F) \equiv (O \vee \exists \vec{v}~H)$ and $G \equiv \neg (F \vee O)$ and $H \wedge O \equiv \textsf{false}$ and $O$ does not mention variables from $\vec{v}$]{$H$ is satisfiable (call \textsc{Smt})}
  \STATE $a \gets$ a model of $H$ \COMMENT{$a \models H$}
  \STATE $M_1 \gets \textsc{Generalize1}(F, a)$ \COMMENT{$M_1 \Rrightarrow F$}
  \STATE $M_2 \gets \textsc{Generalize2}(G, M_1)$ \COMMENT{$\neg(M_2 \wedge G)$}
  \STATE $\pi \gets \textsc{Project}(M_2, \vec{v})$
    \COMMENT{$\pi \equiv \exists \vec{v}~M_2$}
  \STATE $O \gets O \vee \pi$
  \STATE $H \gets H \wedge \neg \pi$
  \STATE $G \gets G \wedge \neg \pi$
\ENDWHILE
\ENSURE{$O \equiv \exists \vec{v}~F$}
\end{algorithmic}
\end{algorithm}

\label{part:algo-mod2}
The algorithm given previously was not the first we experimented; we had originally a slightly more complicated one, given as \textsc{ExistElim}(Mod2), which we wrongly thought would be more efficient. Instead of using $\neg F$ to check for inappropriate generalizations, we used a formula $G$ initially equal to $\neg F$, and then progressively altered. The termination proof stays the same, while correctness relies on the additional invariant $G \equiv \neg(F \vee O)$. \textsc{ExistElim} can be thought of as identical to  \textsc{ExistElim}(Mod2) except that $G$ stays constant.

We thought this scheme allowed more generalization of models than the algorithm we gave earlier in the article, as shown by Fig.~\ref{fig:generalize2}. \textsc{ExistElim} tries to generalize $M$ to a conjunction that implies $F$, but in fact this is too strict a condition to succeed, whereas \textsc{ExistElim}(Mod2) succeeds in generalizing $F$ to a conjunction that implies $F \vee O$. If at least one variable is projected out, and $F$ actually depends on that variable, then the models of $F$ are strictly included in those of the final value of $O$, which is equivalent to $\exists \vec{v}~F$.

Experiments (\S\ref{part:benchmarks}) however showed that this ``more clever'' algorithm is slower by approximately a factor of two, because adding extra assertions to $G$ is costly for the SMT-solver.

\subsection{Extra Modulo Theory}
The algorithm can be easily extended to quantifier elimination modulo an assumption~$T$ on the free variables of~$F$. All definitions stay the same except that $\Rrightarrow$ is replaced by $\Rrightarrow_T$, defined as $P \Rrightarrow_T Q \definedAs (P \wedge T) \Rrightarrow (Q \wedge T)$ and $\equiv$ is replaced by $\equiv_T$, defined as $(P \equiv_T Q) \definedAs (P \wedge T \equiv Q \wedge T)$. \textsc{ExistElim} is modified by replacing the initialization of $G$ and $H$ by $\neg F \wedge T$ and $F \wedge T$ respectively. Intuitively, $T$ defines a universe of validity such that values outside of the models $T$ are irrelevant to the problem being studied.

\section{Comparison with Other Algorithms}
The ``classical'' algorithm for quantifier elimination over linear inequalities is Ferrante and Rackoff's~\cite{FerranteRackoff75}. Another algorithm based on similar ideas, but with better performance, was proposed by Loos and Weispfenning \cite{LoosWeispfenning93}.
We shall therefore compare our method to these algorithms, both theoretically and experimentally. We also compared our algorithm with other available packages using other quantifier elimination techniques.

\subsection{Complexity bounds}
\label{part:complexity}
\begin{table}[htb]%
\renewcommand{\footnoterule}{}% no ruler before footnotes
\makeatletter%
\setlength{\skip\@mpfootins}{1ex}% reduce space between table and footnotes
\makeatother%
\begin{minipage}{\textwidth}%
\begin{center}\small%
\begin{tabular}{|l|r|r|r|r|}
\hline
Benchmark & r. lim. $\bbR$ & r. lim. float & \texttt{prsb23} & \texttt{blowup5} \\
\hline
\textsc{Mjollnir} & 1.4 & 17 & 0.06 & negligible \\
\textsc{Mjollnir} (mod1) & 1.6 & 77%
\footnote{Memory consumption grows to 1.1~GiB.} & 0.06 & negligible\\
\textsc{Mjollnir} (mod2) & 1.5 & 34 & 0.07 & negligible\\
\textsc{Mjollnir} Loos-Weispfenning & o-o-m & o-o-m & o-o-m & negligible \\
Proof-of-concept & n/a & 823 & n/a & n/a \\
\textsc{Mjollnir} Ferrante-Rackoff & o-o-m & o-o-m & o-o-m & negligible \\
Proof-of-concept & n/a & 823 & n/a & n/a \\
\textsc{Lira} & o-o-m & o-o-m & 8.1 & 0.6\\
\textsc{Redlog} \texttt{rlqe} & 182  & o-o-m & 1.4 & negligible \\
\textsc{Redlog} \texttt{rlqe}+\texttt{rldnf} & o-o-m  & o-o-m  & n/a & n/a\\
\textsc{Mathematica} \texttt{Reduce} & (> 12000) & o-o-m & (> 780) & 7.36\\
\hline
\end{tabular}
\end{center}
\end{minipage}

\caption{Timings (in seconds, on an AMD Turion TL-58 64-bit Linux system) for eliminating quantifiers from our benchmarks. The first line is the algorithm described in this paper, the two following linear variants from \S\ref{part:variants}, then other packages. \textsc{Reduce} has \texttt{rlqe} (quantifier elimination) and \texttt{rlqe}+\texttt{rldnf} (same, followed by conversion to DNF). $(> t)$ means that the computation was killed after $t$ seconds because it was running too long. The \texttt{prsb23} and following are decision problems, the output is \texttt{true} or \texttt{false}, thus DNF form does not matter. Out-of-memory is noted ``o-o-m''.}
\label{tab:benchmarks1}
\end{table}

We consider in this section that inequalities are written using integer coefficients in binary notation. We shall prove that a complexity bound $2^{n^{2^q}}$ where $n$ is the number of atomic formulas and $q$ is the number of quantifiers to be eliminated. This yields an overall complexity of $2^{2^{2^{|F|}}}$ where $|F|$ is the size of the formula.

Let us consider a conjunction of inequalities taken from a set of $n$ inequalities. The Fourier-Motzkin algorithm \cite{BradleyManna07,imbert93fouriers} eliminates variable $x$ from this conjunction as follows. It first partitions these inequalities into those where $x$ does not appear, which are retained verbatim, and those where $x$ appears positively ($E_+$) and negatively ($E_-$). From each couple of inequalities $(e_+, e_-)$ in $E_+ \times E_-$, an inequality where $x$ does not appear is obtained by cancellation between $e_+$ and $e_-$. The size in bits of the coefficients in the output inequalities can be at most $2s+1$ where $s$ is the maximal size of the input coefficients.

The inequalities output therefore belong to a set of size asymptotically at most $n^2/4$ (the worst-case occurs when the inequalities split evenly between those in which $x$ appears positively and those where it appears negatively). The output conjunction is in general too large: many inequalities in it are superfluous; yet it is guaranteed to include all inequalities defining the facets of the projection of the polyhedron.

Consider a formula $F$ written with inequalities $A_1,\dots,A_n$ as atomic formulas, with maximal coefficient size $s$. Our algorithm eliminates the quantifier from $\exists x~F$ and outputs a DNF formula $F'$ built with inequalities found in the output of the Fourier-Motzkin algorithm operating on the set $A_1,\dots,A_n$ and variable~$x$. It follows that $F'$ is built from at most, asymptotically, $n^2/4$ inequalities as atomic formulas. The running time for this quantifier elimination comes from:
\begin{itemize}
\item The SMT solving passes. There are at most $2^n$ branches to explore in total. For each branch, SMT has to test whether the solution set of a conjunction of polynomial inequalities is empty or not, which is a particular case of linear programming, with polynomial complexity. The overall SMT cost is therefore bounded by $O(2^n.P(n))$ for some polynomial $P$;
\item The projections, with complexity $O(n^2.s)$, applied to each of at most $2^n$ polyhedra.
\end{itemize}
This gives an overall complexity of $O(2^{cn})$ where $c$ is a constant.

Consider now a succession of quantifier eliminations (with or without alternations). We now have $F$ consisting of a sequence of quantifiers followed by a quantifier-free formula built out of atomic formulas $A_1,\dots,A_n$. Our algorithm performs eliminations in sequence, starting from the rightmost quantifier.

Let us note $A^{(k)}$ the set of atomic formulas that can be obtained after $k$ eliminations; $A^{(0)}=\{A_1,\dots,A_n\}$. Clearly, $|A^{(k)}| \leq |A^{(0)}|^{2^k}$ asymptotically, since at each iteration the size of the set of atomic formulas can at most get squared by Fourier-Motzkin elimination. The size of the coefficients grows at most as $s.2^k$. This yields the promised bound.

It is possible that the bound $|A^{(k)}| \leq |A^{(0)}|^{2^k}$, obtained by observation of the Fourier-Motzkin algorithm, is too pessimistic. The literature does not show examples of such doubly exponential blowups, while polyhedra with single exponential blowups can be constructed.

% \begin{lemma}
% There exist a family of polyhedra in tridimensional space, indexed by $n$, with $n+3$ facets such that their projection onto a plane has $2n+2$ facets.~\cite{Knox_polytopes_PhD_1996}
% \end{lemma}

% \begin{proof}
% The construction is shown in Fig.~\ref{fig:projection_blowup} for $n=4$. One first draws a $n+1$ edge convex polygon in the $(x,z)$ plane (shown in bold at bottom of the figure: edges $H$, $1, \dots, n$), including a $z=0$ edge and other edges more or less in a semicircular shape (one can for instance take half a $2n$-edge regular polygon). This polygon is extruded along the $y$ axis, and the resulting infinite prism is cut at at angle at both ends (the bold lines $A$ and $B$ at the right of the figure), thus producing a polyhedron with $n+3$ facets. Its projection on the $(x,y)$ plane has $2n+2$ facets, as shown on the figure.
% \end{proof}

% \begin{corollary}\label{cor:projection_exponential}
% For any $0 < \alpha < 2$ there exists a family of polyhedra indexed by $k$ in $3k$-dimensional space, with a number $f_k$ of facets, such that there exists a choice of dimensions such that their projection on $2k$-dimensional space has more than $f_k \alpha^k$ facets.
% \end{corollary}

% \begin{proof}
% In a $3k$-dimensional space, consider the product of $k$ polyhedra of tridimensional space with $n+3$ facets as produced by the preceding lemma. This product is a polyhedron with $(n+3)^k$ facets. Its projection, leaving one every third dimension, has $(2n+2)^k$ facets. With $n$ large enough, $\frac{(2n+2)^k}{(n+3)^k} > \alpha^k$.
% \end{proof}

The ``classical'' algorithm for quantifier elimination over real or rational arithmetic is Ferrante and Rackoff's method \cite{FerranteRackoff75}\cite[\S 7.3]{BradleyManna07}\cite[\S 4.2]{NipkowIJCAR08}. A related algorithm was proposed by Loos and Weispfenning \cite{LoosWeispfenning93}\cite[\S 4.4]{NipkowIJCAR08}. Both these algorithms are based on the idea that an existentially quantified formula $\exists x~F(x)$ with free variables $y, z, \dots$ can be replaced by $F(x_1) \vee \dots \vee F(x_m)$ where $x_1,\dots,x_m$ are expressed as functions of $y,z,\dots$. In the case of Ferrante and Rackoff, $m$ is quadratic in the worst case in the length of the formula, while for Loos and Weispfenning it is linear. In both cases, the overall complexity bound is $2^{2^{cn}}$.

The weakness of both algorithms is that they never simplify formulas. This may explain that while their theoretical bounds are better than ours, our algorithm is in practice more efficient, as shown in the next subsection.

One could at first assume that the complexity bounds for our algorithm are asymptotically worse than Ferrante and Rackoff's. Our algorithm, however, outputs results in CNF or DNF form, while Ferrante and Rackoff's algorithm does not. If we add a step of transformation to CNF or DNF to their algorithm, then we also obtain a triple exponential bound.

\subsection{Practical results}
\label{part:benchmarks}

\begin{table}[htb]
\begin{center}
\tableaularge
\begin{tabular}{|l|r|r|r|r|r|r|r|r|r|}
\hline
& \multicolumn{3}{|c}{depth 14}
& \multicolumn{3}{|c|}{depth 15}
& \multicolumn{3}{|c|}{depth 16}\\
& Solved & Avg & O-o-m
& Solved & Avg & O-o-m
& Solved & Avg & O-o-m\\
\hline
\textsc{Mjollnir} & 100 & 1.6 & 0 & 94 & 9.8 & 0 & 73 & 35.3 & 0\\
\textsc{Mjollnir} (mod1) & 94 & 8.2 & 3 & 80 & 27.3 & 7 & 39 & 67.1 & 25\\
\textsc{Mjollnir} (mod2) & 100 & 3.8 & 0 & 91 & 13.9 & 0 & 65 & 39.2 & 0\\
\textsc{Mjollnir} Loos-W. & 93 & 1.77 & 4 & 90 & 6.42 & 5 & 62 & 17.65 & 27\\
Proof-of-concept & 94 & 1.4 & 0 & 86 & 2.2 & 0 & 55 & 17.7 & 0\\
\textsc{Mjollnir} Ferrante-R. & 51 & 18.2 & 41 & 23 & 23.2 & 65 & 3 & 7.3 & 85\\
Proof-of-concept & 94 & 1.4 & 0 & 86 & 2.2 & 0 & 55 & 17.7 & 0\\
\textsc{Lira} & 14 & 102.4 & 83 & 3 & 77.8 & 94 & 1 & 8 & 95\\
\textsc{Redlog} (\texttt{rlqe}) & 92 & 13.7 & 0 & 53 & 27.4 & 0 & 27 & 33.5 & 0\\
\textsc{Mathematica} & 6 & 30.2 & 0 & 1 & 255.7 & 0 & 1 & 19.1 & 0\\
\hline
\end{tabular}
\end{center}
\caption{Benchmarks on $3\times 100$ random instances generated using \texttt{randprsb}, with formula depths $n$ respectively 14, 15 and 16 (obtained by\texttt{randprsb 0 7 -10 10 $n$ $i$}) where $i$ ranges in $[0,99]$). The table shows the number of instances solved within the timeout period out of the proposed 100, the average time spent per solved instance, and the number of instances resulting in out-of-memory.}
\label{tab:benchmarks2}
\end{table}

We benchmarked several variants of our method against other algorithms:
\begin{description}
\item[Mjollnir] is the algorithm described in \S\ref{part:algorithm}, implemented on top of SMT solver \textsc{Yices}\footnote{\url{http://yices.csl.sri.com/}} and the \textsc{NewPolka} polyhedron package from \textsc{Apron}\footnote{\url{http://apron.cri.ensmp.fr/library/}}, or optionally the Parma Polyhedra Library (PPL\footnote{\url{http://www.cs.unipr.it/ppl/}}). Profiling showed that most of the time is spent in the SMT solver, so performance differences between NewPolka and PPL are negligible.

\item[Proof-of-concept] is an early version of the same algorithm, implemented on top of a rudimentary SMT solver and the PPL. The SMT algorithm used is simple and lazy: the SMT problem is turned into SAT by replacing each atomic inequality by a propositional variable, and the SAT problem is input into \textsc{Minisat}. A full SAT solution is obtained, then tested for emptiness by solving a linear programming problem: finding a vector of coefficients suitable as a contradiction witness for Farkas' lemma. If a witness is found, it yields a contradictory conjunction, whose negation is added to the SAT problem and SAT is restarted.

\item[Mjollnir (mod1)] is the ALL-SAT then projection algorithm from~\S\ref{part:algo-mod1}. It is invoked by option \texttt{--no-block-projected-model}.

\item[Mjollnir (mod2)] is the algorithm from \S\ref{part:algo-mod2}; it is invoked by option \texttt{--add-blocking-to-g}.

\item[Mjollnir Ferrante-Rackoff] implements \cite{FerranteRackoff75}\cite[\S 7.3]{BradleyManna07}.

\item[Mjollnir Loos-Weispfenning] implements \cite{LoosWeispfenning93}.

\item[Lira\footnotemark]\footnotetext{\url{http://lira.gforge.avacs.org/}} is based on B\"uchi automata and handles both Presburger arithmetic (integer linear inequalities) and rational linear inequalities.

\item[Mathematica\footnotemark]\footnotetext{\url{http://www.wolfram.com/}} is a general-purpose symbolic algebra package. Its \texttt{Reduce} fonction appears to implement CAD~\cite{CAD_1998}, an algorithm suitable for nonlinear inequalities interpreted in the theory of real closed fields, though it is difficult to know what exactly is implemented because this program is closed source.

\item[Redlog\footnotemark]\footnotetext{\url{http://www.algebra.fim.uni-passau.de/~redlog/}} is a symbolic formula package implemented on top of the computer algebra system \textsc{Reduce}~3.8.\footnote{\url{http://www.uni-koeln.de/REDUCE/}} \textsc{Redlog} implements various algorithms due to Volker Weispfenning and his group~\cite{LW93}.
\end{description}

Table~\ref{tab:benchmarks1} compares these various implementations on a few benchmark examples
\footnote{Available from \url{http://www-verimag.imag.fr/~monniaux/download/linear_qe_benchmarks.zip}.} coming from two sources:
\begin{enumerate}
\item Examples produced from problems of program analysis following our method for the parametric computation of least invariants.~\cite{Monniaux_SAS07}
To summarize, each formula expresses the fact that a set of program states (such as a product of intervals for the numerical variables) is the least invariant of a program, or the strongest postcondition if there is no fixed point involved. Most of the examples, being extracted by hand from simple subprograms, were easily solved and thus did not constitute good benchmarks, but one of them, defining the least invariant of a rate limiter, proved to be tougher to solve, and we selected it as a benchmark. We have two versions of this example: the first for a rate limiter operating over real numbers (``r.~lim $\bbR$'')
the second over floating-point numbers, abstracted using real numbers (``r.~lim float''), and considerably tougher to process than the real example.
\item Examples procured from the \textsc{Lira} designers (\texttt{prsb23} and \texttt{blowup5}).
\end{enumerate}

Memory consumption stayed modest for all examples (< 15~MiB), except for r.~lim float.
Profiling showed that most of the time is spent in the SMT-solver and only a few percents in the projection algorithm. The fact that the proof-of-concept implementation, with a very naive SMT-solver, performs decently on an example where other algorithms exhaust memory shows that the performance of our algorithm cannot be solely explained by the good quality of \textsc{Yices}.

Table~\ref{tab:benchmarks2} compares the various algorithms on random examples.
We then used the LIRA team's \texttt{randprsb} tool%
\footnote{\url{http://lira.gforge.avacs.org/toolpaper/randPrsb.hs}}
to generate $100$ random instances, by changing the seed of the random
number generator from 0 to~99, for each of three values (14, 15, 16) of the depth parameter, which measures complexity.%
\footnote{We used the command line \texttt{randprsb 0 7 -10 10 $n$ $i$} where $n$ is the depth parameter (here, 14, 15 or 16) and $i$ ranges in $[0,99]$.}
The programs were then tested with both
a 1.8~GiB memory limit and a timeout of five minutes.
It is clear from Tab.~\ref{tab:benchmarks2} that
\texttt{Mjollnir --no-add-blocking-to-g} is the most efficient of the
tested tools.

\section{Conclusion and Future Work}
We have proposed a new quantifier elimination algorithm for the theory of linear inequalities over the real or rational numbers, and investigated possible variants. Our motivation was the practical application of a recent result of ours on program analysis, stating that formulas for computing the least invariants of certain kinds of systems can be obtained through quantifier elimination~\cite{Monniaux_SAS07}.

This algorithm is efficient on examples obtained from this program analysis technique, as well as other examples, whereas earlier published algorithms, as well as several commercial packages, all exhaust time or memory resources. Our algorithm leverages the recent progresses on satisfiability modulo theory solvers (SMT) and, contrary to older algorithms, performs on-the-fly simplifications of formulas that keep formula sizes manageable. Our algorithm also performs better than a straight application of SMT solvers (ALL-SAT followed by projection).

Our algorithm is described for rational or real linear arithmetic, but it can be extended to any theory for which there is an efficient satisfiability testing algorithm for unquantified formulas and a reasonably efficient projection algorithm for conjunctions. Among extensions that could be interesting from a practical point of view would be on the one hand the nonlinear case for real arithmetic (polynomials), and on the other hand the mixed integer / real problems. Of course, nonlinear integer arithmetic cannot be considered, since Peano arithmetic is undecidable.

Tarski showed that the theory of the real closed fields (inequalities of polynomial expressions) admits quantifier elimination,~\cite{Tarski51} however his algorithm had impractical (non-elementary) complexity. Later, the \emph{cylindrical algebraic decomposition} (CAD)~\cite[Ch.~11]{Basu_Pollack_Roy_2003} method was introduced, with doubly exponential complexity, which is unavoidable in the worst case~\cite[\S 11.4]{Basu_Pollack_Roy_2003}. Our experiments with both \textsc{Mathematica} and \textsc{Qepcad}, both of which implement CAD, as well as with \textsc{Reduce}/\textsc{Redlog}, which implement various algorithms for quantifier elimination, showed us that combinatorial blowup occurs very quickly. For such techniques to be interesting in practice, practical complexity should be lowered. Perhaps our technique could help. There are, however, significant difficulties in that respect. Our technique starts with some single model of the target formula over the rational numbers; but a system of nonlinear inequalities needs not have rational models when it is not full-dimensional (for instance, $X^2=2$). Our technique reduces the geometrical computations to computations on conjunctions; but in the nonlinear case, single inequalities can be reduced to disjunctions. As an example, $X^2 \geq 4$ is reduced to $X \leq -2 \vee X \geq 2$. Most importantly, our technique relies at several steps on the availability of a decision procedure that stays efficient even when the answer is negative.

Regarding the mixed integer / real problems, the \textsc{Lira} tool implements quantifier elimination using a weak form of B\"uchi automata matching the $b$-ary expression of the integers or reals, where $b$ is an arbitrary base.~\cite{Becker_et_al_CAV05} The output of the process  is an automaton and not a readable formula. While it is possible to decide a closed formula, and to obtain one model from a satisfiable non-closed formula, it is an open problem how to efficiently reconstruct a quantifier-free formula from the resulting automaton. The automaton construct is unsuitable for large coefficients (as our examples obtained from the analysis of floating-point programs). Even on examples with small coefficients, the tool was unable to complete quantifier elimination without blowing up. We think therefore that it would be interesting to be able to apply our technique to the mixed integer / real problems, but there are difficulties: the algorithms on integer polyhedra are considerably more complex than on rational polyhedra.

A classical objection to automatic program analysis tools meant to prove the absence of bugs is that these tools could themselves contain bugs. Our method uses complex algorithms (SMT-solving, polyhedron projection) as sub-procedures. We consider developing techniques so that the algorithm outputs easily-checkable proofs or ``proof witnesses'' of the correctness of its computation. Furthermore, we showed in earlier publications \cite{Monniaux_SAS07} that certain program analysis tasks were equivalent to quantifier elimination problems; that is, an effective static analyzer can be extracted from the quantifier-free form of an analyzer specification. This therefore suggests a new way for writing safe static analyzers: instead of painstakingly writing an analyzer, then proofs of correctness in a proof assistant \cite{Pich:these}, one could formulate the analysis problem as an equivalent quantifier elimination problem, with a relatively simple proof of equivalence, then apply a ``certified'' quantifier elimination procedure in order to extract the effective analyzer.

\bibliography{monniaux-lpar08}